\documentclass[aps,prd,preprint,preprintnumbers,nofootinbib,superscriptaddress]{revtex4}
\usepackage{ae}
\usepackage{bm} 
\usepackage{color}
\usepackage{amsmath}
\usepackage{amssymb}
\usepackage{amsfonts}
\usepackage{graphicx}
\usepackage{slashed}
\usepackage{setspace}
\usepackage{stackrel}
\usepackage{ulem}

\newcommand{\Eqref}[1]{Eq.~\eqref{#1}}

\allowdisplaybreaks

\begin{document}

\setlength{\unitlength}{1mm}
\title{Tadpole diagrams in constant electromagnetic fields}
\author{Felix Karbstein}\email{felix.karbstein@uni-jena.de}
\affiliation{Helmholtz-Institut Jena, Fr\"obelstieg 3, 07743 Jena, Germany}
\affiliation{Theoretisch-Physikalisches Institut, Abbe Center of Photonics, \\ Friedrich-Schiller-Universit\"at Jena, Max-Wien-Platz 1, 07743 Jena, Germany}

\date{\today}

\begin{abstract}
 We show that all possible one-particle reducible tadpole diagrams in constant electromagnetic fields can be constructed from one-particle irreducible constant-field diagrams.
 The construction procedure is essentially algebraic and involves differentiations of the latter class of diagrams with respect to the field strength tensor and contractions with derivatives of the one-particle irreducible part of the Heisenberg-Euler effective Lagrangian in constant fields.
 Specific examples include the two-loop addendum to the Heisenberg-Euler effective action as well as a novel one-loop correction to the charged particle propagator in constant electromagnetic fields discovered recently.
 As an additional example, the approach devised in the present article is adopted to derive the tadpole contribution to the two-loop photon polarization tensor in constant fields for the first time.
\end{abstract}

\maketitle

\section{Introduction}

In Ref.~\cite{Gies:2016yaa} it has recently been emphasized that by definition the Heisenberg-Euler effective action $\Gamma_\text{HE}$ \cite{Heisenberg:1935qt,Weisskopf} is not one-particle irreducible (1PI) if relying on Schwinger's definition \cite{Schwinger:1951nm}.
One-particle reducible (1PR) tadpole diagrams, i.e., Feynman subdiagrams consisting of quantum fluctuations of charged particles that couple to the rest of the diagram only via a single photon line, start contributing at two loops.
As constant fields cannot supply momentum to virtual particle loops, momentum conservation requires the latter photon line to transfer zero momentum.  
The simplest tadpole diagram consists of a single charged-particle loop, dressed in the constant field, which couples to the rest of the diagram via a photon line.
Remarkably, these contributions are nonvanishing even in constant electromagnetic fields, and give rise to novel contributions previously believed to vanish.

The first explicit result for such a contribution was provided in Ref.~\cite{Gies:2016yaa}, who evaluated the 1PR contribution to $\Gamma_\text{HE}$ in constant fields.
Similar contributions are expected to become relevant in the context of higher-loop calculations of essentially all physical quantities to be studied in external electromagnetic fields. 
In the meantime, Refs.~\cite{Edwards:2017bte,Ahmadiniaz:2017rrk} have evaluated the corresponding tadpole contributions to the propagators of charged spinor and scalar particles in constant fields at one loop level.

In this article, we present a straightforward and efficient approach to explicitly determine such 1PR tadpole contributions for generic quantities evaluated in constant external fields.
Our strategy is heavily based on the fact that no momentum can be transferred from tadpole diagrams evaluated in constant fields.
The main result of the present study is the observation that all 1PR tadpole diagrams contributing to a given quantity in constant fields can be generated by differentiations of lower loop diagrams for the field strength tensor and contractions with derivatives of the Heisenberg-Euler effective Lagrangian. Along these lines, we demonstrate that all possible 1PR tadpole diagrams in constant fields can be generated recursively by differentiations of 1PI constant-field diagrams.
Hence, even though 1PR tadpole contributions are relevant for the physics in constant fields, all the information required for their explicit determination is encoded in 1PI building blocks.

As an exemplary application of our considerations, we derive the exact expression for the 1PR contribution to the two-loop photon polarization tensor in generic constant electromagnetic fields.
Here, our main focus is on a purely magnetic background field. However, the obtained results can be directly translated to a purely electric field by means of an electric-magnetic duality.

Our article is organized as follows. In Sec.~\ref{sec:SofA}, we recall the state of the art for calculations of 1PR tadpole contributions in constant external fields.
So far, exact results are available for the Heisenberg-Euler effective Lagrangian (Sec.~\ref{subsec:SofA1}) and for the propagators of scalar and spinor charged particles (Sec.~\ref{subsec:SofA2}).
Section~\ref{sec:1PIto1PR} details on how the 1PR tadpole contributions to a given quantity evaluated in constant fields can be generated from its 1PI sector.
In Sec.~\ref{sec:2loopPi} we exemplarily derive the tadpole contribution to the two-loop photon polarization tensor in constant fields. After discussing the case of generic constant electromagnetic background fields (Sec.~\ref{subsec:2loopPi1}), we focus on a purely magnetic field (Sec.~\ref{subsec:2loopPi2}). 
Finally, we end with conclusions and a brief outlook in Sec.~\ref{sec:CaO}.

\section{State of the art}\label{sec:SofA}

\subsection{Heisenberg-Euler effective Lagrangian}\label{subsec:SofA1}

In this section, we briefly recall the determination of the 1PR contribution to the Heisenberg-Euler effective Lagrangian at two loops \cite{Gies:2016yaa}, specializing to a constant backgrounds field from the outset.
In $d$ space-time dimensions and in constant electromagnetic fields $F\equiv F^{\mu\nu}$, we have $\Gamma_\text{HE}=V^{(d)}{\cal L}_\text{HE}$, where $V^{(d)}$ denotes the space-time volume and ${\cal L}_\text{HE}\equiv{\cal L}_\text{HE}(F)$ is the Heisenberg-Euler effective Lagrangian evaluated in the prescribed field. Due to the fact that $\Gamma_\text{HE}$ differs from ${\cal L}_\text{HE}$ just by an overall volume factor, both quantities are trivially related in constant fields.

\begin{figure}[h]
 \centering
 \includegraphics[width=0.45\columnwidth]{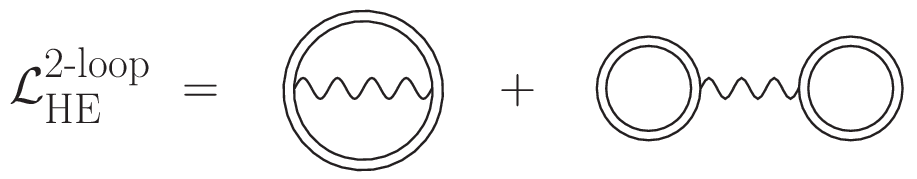}
\caption{Feynman diagrams constituting ${\cal L}^{2\text{-loop}}_\text{HE}={\cal L}^{2\text{-loop}}_{1\text{PI}}+{\cal L}^{2\text{-loop}}_{1\text{PR}}$. The double solid line denotes the fermion propagator dressed to all orders in the external electromagnetic field.}
\label{fig:Fig1}
\end{figure}

The Heisenberg-Euler effective Lagrangian naturally decomposes into 1PI and 1PR contributions, and can be expanded in terms of loops, i.e.,
\begin{equation}
 {\cal L}_\text{HE}={\cal L}_{1\text{PI}}+{\cal L}_{1\text{PR}}=\sum_{\ell=1}^\infty{\cal L}_\text{HE}^{\ell\text{-loop}}\,,
\end{equation}
with ${\cal L}^{\ell\text{-loop}}_\text{HE}\sim(\frac{\alpha}{\pi})^{\ell-1}$, where $\alpha=\frac{e^2}{4\pi}\simeq\frac{1}{137}$ is the fine-structure constant; we use the Heaviside-Lorentz System with
$c=\hbar=1$, and $e$ is the elementary charge. 
Correspondingly, analogous expansions exist for ${\cal L}_{1\text{PI/PR}}$, and we have ${\cal L}_{1\text{PI/PR}}^{\ell\text{-loop}}={\cal L}_\text{HE}^{\ell\text{-loop}}\big|_{1\text{PI}/\text{PR}}$.
At one-loop level, ${\cal L}_\text{HE}$ is determined by a single fermion loop dressed in the external field, such that  ${\cal L}_{1\text{PI}}^{1\text{-loop}}={\cal L}_\text{HE}^{1\text{-loop}}$ and ${\cal L}_{1\text{PR}}^{1\text{-loop}}=0$.
So far, only ${\cal L}_\text{HE}^{1\text{-loop}}$ and ${\cal L}_\text{HE}^{2\text{-loop}}$ are known explicitly in $d=3+1$ dimensions \cite{Heisenberg:1935qt,Weisskopf,Schwinger:1951nm,Ritus:1975,Ritus:1977,Dittrich:1985yb,Fliegner:1997ra,Kors:1998ew,Gies:2016yaa}; cf. Ref.~\cite{Dunne:2004nc} for a review.
On the three-loop level, first analytical results for ${\cal L}_{1\text{PI}}^{3\text{-loop}}$ have been obtained in $d=1+1$ dimensions \cite{Huet:2009cy,Huet:2011kd}.
For a graphical representation of ${\cal L}^{2\text{-loop}}_\text{HE}$ for quantum electrodynamics (QED) in terms of Feynman diagrams, see Fig.~\ref{fig:Fig1}.

The 1PR addendum at two loops (right diagram in Fig.~\ref{fig:Fig1}) has first been evaluated by Ref.~\cite{Gies:2016yaa}, who showed that it is fully determined by ${\cal L}^{1\text{-loop}}_\text{HE}$ and reads
\begin{equation}
  {\cal L}^{2\text{-loop}}_{1\text{PR}}=\frac{2}{d}\frac{\partial{\cal L}^{1\text{-loop}}_\text{HE}}{\partial F^{\mu\nu}}\frac{\partial{\cal L}^{1\text{-loop}}_\text{HE}}{\partial F_{\mu\nu}}\,.
 \label{eq:1PR@2loop}
\end{equation}
This expression is obtained upon contraction of two one-loop photon currents $j^{1\text{-loop}}_\mu$ with the photon propagator,
\begin{equation}
  D^{\mu\nu}(k)=\frac{1}{k^2-{\rm i}\epsilon}\Bigl(g^{\mu\nu}-(1-\xi)\frac{k^\mu k^\nu}{k^2-{\rm i}\epsilon}\Bigr)\,,
\end{equation}
where $\xi=1$ in Feynman gauge, and integrating over the momentum transfer $k$; $\epsilon\to0^+$.
More specifically, the formal expression corresponding to the handcuff Feynman diagram depicted in Fig.~\ref{fig:Fig1} reads
\begin{equation}
 {\cal L}^{2\text{-loop}}_{1\text{PR}}=\frac{1}{V^{(d)}}\frac{1}{2}\int\frac{{\rm d}^dk}{(2\pi)^d}\,j_\mu^{1\text{-loop}}(k)D^{\mu\nu}(k)j_\nu^{1\text{-loop}}(-k)\,.
 \label{eq:1PR@jDj}
\end{equation}

In momentum space, the 1PI photon current in constant electromagnetic fields is given by \cite{Gies:2016yaa}
\begin{equation}
  j_\mu(k)=2{\rm i}\,(2\pi)^d\delta^{(d)}(k)\biggl[\frac{\partial{\cal L}_{1\text{PI}}}{\partial F^{\alpha\mu}}\,k^\alpha\,+{\cal O}(k^3)\biggr]\,.
  \label{eq:jnloop}
\end{equation}
Here, the overall delta function in $d$ space-time dimensions reflects the fact that a constant field cannot supply momentum to virtual charged particle fluctuations,
while the linear dependence of $k^\alpha$ is a direct consequence of the Ward identity ensuring $k^\mu j_\mu(k)=0$.
Let us emphasize, that the current~\eqref{eq:jnloop} encompasses all possible 1PI contributions of virtual particle fluctuations arising in QED in constant external fields, which exhibit a single coupling to the photon field.
For QED at zero field, this current vanishes because of Furry's theorem, forbidding non-zero contributions from fermion loops with an odd number of photon couplings.
Upon limitation to the one-loop contribution of \Eqref{eq:jnloop}, which amounts to substituting ${\cal L}_{1\text{PI}}\to{\cal L}^{1\text{-loop}}_\text{HE}$, and making use of the fact that
\begin{equation}
 \int\frac{{\rm d}^d k}{(2\pi)^d} \frac{k^\alpha k^\beta}{k^2-{\rm i}\epsilon}\,\bigl[(2\pi)^d\delta^{(d)}(k)\bigr]^2
 =\frac{g^{\alpha\beta}}{d}\int\frac{{\rm d}^dk}{(2\pi)^d} \bigl[(2\pi)^d\delta^{(d)}(k)\bigr]^2
 =\frac{g^{\alpha\beta}}{d}V^{(d)}\,,
\end{equation}
\Eqref{eq:1PR@2loop} follows straightforwardly from \Eqref{eq:1PR@jDj}.

\begin{figure}[h]
 \centering
 \includegraphics[width=0.40\columnwidth]{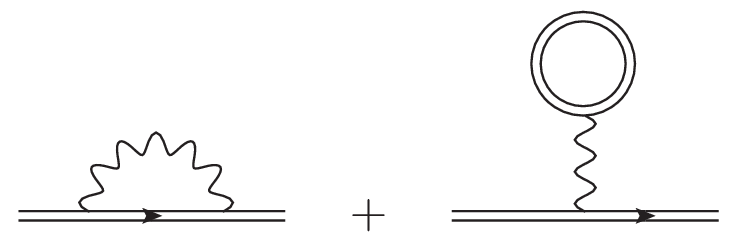}
\caption{One-loop corrections to the charged particle propagator in an external electromagnetic field.}
\label{fig:Fig2}
\end{figure}

Prior to the recent analysis performed in Ref.~\cite{Gies:2016yaa}, this contribution was believed to vanish in constant fields \cite{Ritus:1975,Dittrich:1985yb}, based on the argument that $(2\pi)^d\delta^{(d)}(k)\,k^\alpha=0$ in \Eqref{eq:jnloop},
thereby missing the subtlety that, upon convolution with the infrared divergent photon propagator, the current~\eqref{eq:jnloop} might induce finite 1PR tadpole contributions in Feynman diagrams.
Finally, note that ${\cal L}^{2\text{-loop}}_{\text{HE}}$ may alternatively be viewed as a one-loop calculation performed with the charged particle propagator in the constant field already accounting for quantum corrections at one-loop level. The latter is depicted in Fig.~\ref{fig:Fig2}.
Identifying the in- and outgoing charged particle lines in Fig.~\ref{fig:Fig2} to form loops, the diagrams shown in Fig.~\ref{fig:Fig1} are recovered. 

\subsection{Charged particle propagators in constant electromagnetic fields}\label{subsec:SofA2}

Meanwhile, Refs.~\cite{Edwards:2017bte,Ahmadiniaz:2017rrk} have explicitly determined the one-loop tadpole corrections for the propagators of both charged scalar and spinor particles in constant fields within the worldline formalism. The respective Feynman diagram is depicted in Fig.~\ref{fig:Fig2} (right).
Their finding is that in momentum space this 1PR contribution to the charged particle propagator can be expressed similarly to \Eqref{eq:1PR@2loop} as
\begin{equation}
 G^{1\text{-loop}}(k|F)\big|_{1\text{PR}}=\frac{\partial G^{0\text{-loop}}(k|F)}{\partial F^{\mu\nu}}\frac{\partial{\cal L}^{1\text{-loop}}_\text{HE}}{\partial F_{\mu\nu}} \,,
 \label{eq:G_1PR}
\end{equation}
where $G^{0\text{-loop}}(k|F)$ denotes the charged scalar/spinor propagator dressed to all orders in the constant external field $F$ \cite{Fock:1937dy,Schwinger:1951nm}, depicted as double solid line in Figs.~\ref{fig:Fig1} and \ref{fig:Fig2}.

\section{From 1PI to 1PR tadpole diagrams in constant fields}\label{sec:1PIto1PR}

Here, we argue that the structural similarity of Eqs.~\eqref{eq:1PR@2loop} and \eqref{eq:G_1PR} is not a coincidence, but rather a direct consequence of the fact that no momentum can be transferred from the induced photon current~\eqref{eq:jnloop} in constant fields.
We find that this implies that all 1PR tadpole contributions to a given quantity follow by differentiations of lower loop diagrams for the field strength tensor and contractions with derivatives of the constant-field Heisenberg-Euler effective Lagrangian.
Along these lines it can in particular be shown that all 1PR tadpole diagrams to a given quantity in constant fields are fully determined by 1PI diagrams.

To this end, let us first detail on how a given 1PI diagram $g_0(F)$, accounting for the couplings to the constant electromagnetic field to all orders, can be used to determine the exact expression for the set\footnote{This set accounts for all possible insertion possibilities and permutations of the 1PI current.} of 1PR tadpole diagrams contributing to the same quantity $g$ but in addition featuring a single coupling to the 1PI photon current~\eqref{eq:jnloop} in constant fields. We denote the latter diagrams by $g_1(F)$, i.e., the index $i$ of $g_i$ counts the number of current insertions, or equivalently, the numbers of zero-momentum photon lines of a set of diagrams.

The basic idea is to replace couplings to the constant external field of $g_0(F)$ by couplings to a dynamical photon field. The latter will then be used to mediate couplings to the 1PI photon current~\eqref{eq:jnloop}.
This strategy is, e.g., routinely employed when deriving nonlinear vacuum effects such as vacuum birefringence from the Heisenberg-Euler Lagrangian in constant fields \cite{BialynickaBirula:1970vy}, employing a locally constant field approximation (LCFA), or equivalently a zeroth-order derivative expansion, for the probe photon field.
Note however, that by construction, the procedure of determining the effective coupling via an LCFA only allows for trustworthy results in the limit of low photon frequencies and momenta; cf., e.g., \cite{Galtsov:1982,Karbstein:2015cpa}.
It becomes exact in the formal limit of photon fields with zero frequency and momentum, $k\to0$.
As will be detailed below, due to the momentum structure of the 1PI photon current~\eqref{eq:jnloop} this is precisely the limit which will be relevant in the following considerations, thereby allowing us to obtain exact results from an LCFA calculation.

To pursue this strategy, we first split the constant external field $F$ as $F\to F+f(x)$ into a constant part and a weakly varying photon field $f^{\mu\nu}(x)=\partial^\mu a^\nu(x)-\partial^\nu a^\mu(x)$.
In inhomogeneous external fields, physical quantities $g$ are generically functionals of the prescribed field, but become just ordinary functions of $F$ in constant fields. Hence, we have
\begin{align}
 g_0(F)\quad\xrightarrow{\ \text{LCFA}\ }&\quad\ \frac{1}{V^{(d)}}\int{\rm d}^dx\,g_0\bigl(F+f(x)\bigr) \nonumber\\
 &=g_0(F)+\frac{1}{V^{(d)}}\frac{\partial g_0(F)}{\partial F^{\mu\nu}}\int{\rm d}^dx\,f^{\mu\nu}(x) +{\cal O}(f^2)\,. \label{eq:LCFA}
\end{align}
Note, that the right-hand side of \Eqref{eq:LCFA} reduces to $g_0(F)$ for $f(x)\to0$.
As we are interested only in the coupling to a single 1PI photon current~\eqref{eq:jnloop}, we terminate the expansion at linear order in $f(x)$.
Terms of ${\cal O}(f^2)$ give rise to effective multi-photon couplings to $g_0$.
A functional derivative of the LCFA expression on the right-hand side of \Eqref{eq:LCFA} for $a^\mu$ then provides us with the effective single photon coupling of $g_0$, which we denote by $\frac{\delta g_0}{\delta a^\mu}$. In momentum space, we obtain
\begin{equation}
 \frac{\delta g_0}{\delta a^\mu}(k)=\frac{2{\rm i}}{V^{(d)}}\frac{\partial g_0(F)}{\partial F^{\beta\mu}}
 \,(2\pi)^d\delta^{(d)}(k)\,k^\beta\,. \label{eq:diffg0mu}
\end{equation}
The right-hand side of \Eqref{eq:diffg0mu} is proportional to $(2\pi)^d\delta^{(d)}(k)$, reflecting the fact that no momentum is transferred from constant fields.

The set of 1PR tadpole diagrams to $g$ featuring a single coupling to the 1PI photon current~\eqref{eq:jnloop} is then given by
\begin{equation}
 g_{1}(F):=\int\frac{{\rm d}^dk}{(2\pi)^d}\,j_\mu(k)D^{\mu\nu}(k)\frac{\delta g}{\delta a^\nu}(-k)
 =\frac{4}{d}\,(\partial_F{\cal L}_{1\text{PI}})\frac{\partial}{\partial F}g_0(F)\,, \label{eq:g1}
\end{equation}
where we made use of the shorthand notation
\begin{equation}
 (\partial_F{\cal L}_{1\text{PI}})\frac{\partial}{\partial F}:=\frac{\partial{\cal L}_{1\text{PI}}}{\partial F^{\mu\nu}}\frac{\partial}{\partial F_{\mu\nu}}\,.
\end{equation}
As the 1PI photon current supplies zero momentum to $\frac{\delta g_0}{\delta a^\mu}(k)$, the integral over $k$ receives all its contributions from the limit $k\to0$ where the LCFA becomes exact, and \Eqref{eq:g1} constitutes an exact result.  
Note, that upon identification of $j_\mu=j_\mu^{1\text{-loop}}$ and $g_0(F)={\cal L}_\text{HE}^{1\text{-loop}}$ we reproduce \Eqref{eq:1PR@2loop}, while for $j_\mu=j_\mu^{1\text{-loop}}$ and $g_0(F)=G^{0\text{-loop}}(k|F)$ we obtain \Eqref{eq:G_1PR}.

Aiming at the determination of 1PR tadpole diagrams with two current insertions $g_2$, we adopt the procedure detailed above to construct $g_1$ from $g_0$ to $g_1$. 
This results in 
\begin{equation}
 g_{2}(F)=\frac{4}{d}\,(\partial_F{\cal L}_{1\text{PI}})\frac{\partial}{\partial F}g_1(F)
 =\Bigl(\frac{4}{d}\,(\partial_F{\cal L}_{1\text{PI}})\frac{\partial}{\partial F}\Bigr)^2g_0(F)\,,
\end{equation}
where the additional derivative for $F$ of course acts on both $(\partial_F{\cal L}_{1\text{PI}})$ and $\frac{\partial g_0(F)}{\partial F}$.
By further iteration it is straightforward to show that the set of diagrams featuring $n$ tadpoles is generated by
\begin{equation}
 g_{n}(F)=\Bigl(\frac{4}{d}\,(\partial_F{\cal L}_{1\text{PI}})\frac{\partial}{\partial F}\Bigr)^n g_0(F)\,. \label{eq:res1}
\end{equation} 

Hence, the complete set of diagrams constituting a given quantity $g$, accounting for all possible tadpole diagrams, follows from \Eqref{eq:res1} by substituting
\begin{align}
 {\cal L}_{1{\rm PI}}&\to\sum_{l=1}^\infty {\cal L}_{1{\rm PI}}^{l\text{-loop}} \,, \nonumber\\
 g_0(F)&\to\sum_{k=0}^\infty g_0^{k\text{-loop}}(F) \,,
\end{align}
and summing over all possible values of $n\in\mathbb{N}_0$.
Note that this set of diagrams encompasses both 1PI and 1PR tadpole contributions: The iterative construction procedure outlined above generically also dresses 1PI tadpoles already inserted in previous steps with additional 1PI tadpoles, thereby inducing 1PR tadpoles.
Apart from potential symmetry factors that can be readily accounted for, the $\ell$-loop contribution to $g$ can then be expressed as
\begin{multline}
 g^{\ell\text{-loop}}(F)=g_0^{\ell\text{-loop}}(F)+\sum_{k=0}^{\ell-1} \,
 \sum_{l_1m_1+\cdots+l_nm_n=\ell-k} \\
 \times 
 \Bigl(\frac{4}{d}(\partial_F{\cal L}_{1\text{PI}}^{l_1\text{-loop}})\frac{\partial}{\partial F}\Bigr)^{m_1} \cdots
 \Bigl(\frac{4}{d}(\partial_F{\cal L}_{1\text{PI}}^{l_n\text{-loop}})\frac{\partial}{\partial F}\Bigr)^{m_n} g_0^{k\text{-loop}}(F) \,, \label{eq:gellloop}
\end{multline}
where the second sum is taken over all sequences of positive integer indices $\{l_i,m_i,n\}\geq1$, with $i\in\{1,\ldots,n\}$, such that $\sum_{i=1}^n l_im_i=\ell-k$.
Since $g_0(F)$ is 1PI by definition, the right-hand side of \Eqref{eq:gellloop} builds up all 1PR and 1PI structures from 1PI building blocks.

Let us also emphasize that \Eqref{eq:gellloop} yields all tadpole diagrams with the correct multiplicities: For instance, the insertion of two one-loop tadpoles into a fermion
line is generated only once, the insertion of one one-loop and one two-loop tadpole twice.

\section{Two-loop photon polarization tensor}\label{sec:2loopPi}

\subsection{Generic constant electromagnetic fields}\label{subsec:2loopPi1}

\begin{figure}[h]
 \centering
 \includegraphics[width=0.40\columnwidth]{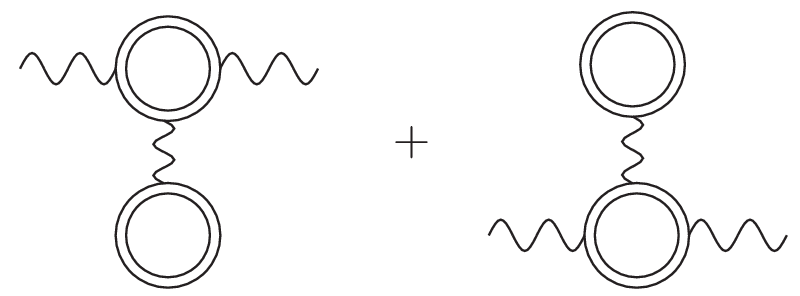}
\caption{Tadpole diagrams contributing to the two-loop photon polarization tensor in constant electromagnetic fields.}
\label{fig:Fig3}
\end{figure}

In the next step, we exemplarily employ the strategy devised in Sec.~\ref{sec:1PIto1PR} to determine the exact expression for the tadpole contribution to the two-loop photon polarization tensor in constant electromagnetic fields for QED in $d=3+1$.
The Feynman diagrams constituting this contribution are shown in Fig.~\ref{fig:Fig3}. Note, that our approach to derive such contributions outlined in the previous section automatically accounts for both depicted diagrams.
Together with the corresponding, not yet determined 1PI contribution, this will constitute the full expression for the photon polarization tensor at two loops.
Due to translational invariance in constant fields, the photon polarization tensor only depends on the momentum transfer $k$, and the ingoing and outgoing momenta agree with each other.
So far, the corresponding photon polarization tensor $\Pi^{\mu\nu}(k|F)$ is only known exactly at one loop level \cite{BatShab,Baier:1974hn,Urrutia:1977xb,Dittrich:2000wz,Schubert:2000yt,Dittrich:2000zu}.
At two loops, first results for both the 1PI and 1PR contributions have been derived only recently \cite{Gies:2016yaa}.
However, due to the fact these findings are based upon an LCFA of the Heisenberg-Euler effective Lagrangian, they are manifestly restricted to the limit of small momentum transfers.

Aiming at determining the 1PR contribution to $\Pi^{\mu\nu}_{2\text{-loop}}(k|F)$ in $d=3+1$, we specialize \Eqref{eq:gellloop} to $g\to\Pi^{\mu\nu}$ and $\ell=2$.
Hence, we have $g_0^{0\text{-loop}}=0$, $g_0^{1\text{-loop}}=\Pi^{\mu\nu}_{1\text{-loop}}(k)$, $g_0^{2\text{-loop}}=\Pi^{\mu\nu}_{2\text{-loop}}(k)\big|_{1{\rm PI}}$, and thus
\begin{equation}
 \Pi^{\mu\nu}_{2\text{-loop}}(k|F)=\Pi^{\mu\nu}_{2\text{-loop}}(k|F)\big|_{1{\rm PI}}
  + \underbrace{(\partial_F{\cal L}_{\text{HE}}^{1\text{-loop}})\frac{\partial}{\partial F} \Pi^{\mu\nu}_{1\text{-loop}}(k|F) }_{=:\Pi^{\mu\nu}_{2\text{-loop}}(k|F)\big|_{1{\rm PR}}} \,.
\end{equation}
For constant external electromagnetic fields, Lorentz and gauge invariance constrain ${\cal L}_\text{HE}$ to depend on $F$ only in terms of the two scalar invariants \cite{Euler:1935zz,Heisenberg:1935qt} ${\cal F}=\frac{1}{4} F_{\mu\nu} F^{\mu\nu}=\frac{1}{2}\bigl(\vec{B}^2-\vec{E}^2\bigr)$ and ${\cal G}=\frac{1}{4} F_{\mu\nu}{}^*\!F^{\mu\nu}=-\vec{B}\cdot\vec{E}$, with dual field strength tensor ${}^*\!F^{\mu\nu}=\frac{1}{2}\epsilon^{\mu\nu\alpha\beta}F_{\alpha\beta}$.\footnote{As $\cal G$ is a pseudoscalar and $\cal L_\text{HE}$ is a scalar quantity, the dependence is actually in terms of ${\cal G}^2$.} Here, $\epsilon^{\mu\nu\alpha\beta}$ is the totally
antisymmetric tensor ($\epsilon^{0123}=1$), and our metric convention
is $g_{\mu \nu}=\mathrm{diag}(-1,+1,+1,+1)$. 
In turn, the 1PR contribution can be expressed as
\begin{equation}
 \Pi^{\mu\nu}_{2\text{-loop}}(k|F)\big|_{1{\rm PR}}
 = \frac{1}{2}\Bigl( \frac{\partial {\cal L}_{\text{HE}}^{1\text{-loop}}}{\partial{\cal F}}F\partial_F
 +\frac{\partial {\cal L}_{\text{HE}}^{1\text{-loop}}}{\partial{\cal G}}{}^*\!F\partial_F \Bigr) \Pi^{\mu\nu}_{1\text{-loop}}(k|F) \,.
 \label{eq:Pi2}
\end{equation}
where we made use of the shorthand notations $F\partial_F:=F^{\rho\sigma}\frac{\partial}{\partial F^{\rho\sigma}}$ and ${}^*\!F\partial_F:={}^*\!F^{\rho\sigma}\frac{\partial}{\partial F^{\rho\sigma}}$.

We have explicitly checked that upon application of this formula to the LCFA result for the one-loop photon polarization tensor in constant fields~\cite{Karbstein:2015cpa}, the low momentum limit of $\Pi^{\mu\nu}_{2\text{-loop}}(k|F)\big|_{1{\rm PR}}$ first obtained in Ref.~\cite{Gies:2016yaa} is correctly reproduced. Reference~\cite{Gies:2016yaa} adopted a slightly different approach to extract this quantity.

However, here our aim is to extract the exact result for $\Pi^{\mu\nu}_{2\text{-loop}}(k|F)\big|_{1{\rm PR}}$.
To this end, we employ the representation of $\Pi^{\mu\nu}_{1\text{-loop}}$ for $d=3+1$ and ${\cal G}\geq0$ given in Ref.~\cite{Dittrich:2000zu},
which makes use of the following decomposition, 
\begin{equation}
    \Pi^{\mu\nu}_{1\text{-loop}}=\Pi_0 P_0^{\mu\nu}+\Pi_\parallel P_\parallel^{\mu\nu} +\Pi_\perp P_\perp^{\mu\nu} +\Pi_Q Q^{\mu\nu} \,,
\end{equation}
where the field dependence of the scalar functions $\Pi_{\cal I}$ with ${\cal I}\in\{0,\parallel,\perp,Q\}$ is via $\cal F$, $\cal G$ and $z_k=(k_\rho F^{\rho\kappa})(k_\sigma F^\sigma_{\,\ \kappa})$ only.
The projectors $P^{\mu\nu}_{0,\parallel,\perp}$ employed here are defined as
\begin{equation}
    P_0^{\mu\nu}=\frac{u^\mu u^\nu}{u^2}\,, \quad
    P^{\mu\nu}_\parallel=\frac{v_\parallel^\mu v_\parallel^\nu}{v_\parallel^2}\,, \quad
    P^{\mu\nu}_\perp=\frac{v_\perp^\mu v_\perp^\nu}{v_\perp^2}\,,
\end{equation}
and the additional tensor structure as $Q^{\mu\nu}=v_\parallel^\mu v_\perp^\nu +v_\perp^\mu v_\parallel^\nu$, with
\begin{align}
 u^\mu&=F^2k^\mu+\frac{z_k}{k^2}k^\mu\,, \nonumber \\
 v_{\parallel/\perp}^\mu&=\frac{1}{c_+^2+c_-^2}\bigl(c_{\pm}{}^*\!Fk^\mu\mp c_\mp Fk^\mu\bigr)\,, \label{eq:vs}
\end{align}
where $F^2k^\mu=F^{\mu\rho}F_{\rho\sigma}k^\sigma$, $c_\pm=(\sqrt{{\cal F}^2+{\cal G}^2}\pm{\cal F})^{1/2}$, ${}^*\!Fk^\mu={}^*\!F^{\mu\rho}k_\rho$ and $Fk^\mu=F^{\mu\rho}k_\rho$.
For completeness, note that the squares of the four-vectors defined in \Eqref{eq:vs} can be expressed as
\begin{align}
 u^2&=2{\cal F}z_k+{\cal G}^2k^2-\Bigl(\frac{z_k}{k^2}\Bigr)^2k^2\,, \nonumber\\
 v_{\parallel/\perp}^2&=\frac{z_k\mp k^2c_\pm^2}{c_+^2+c_-^2}\,.
\end{align}
Finally, the scalar quantities $\Pi_{\cal I}$ for generic constant electromagnetic fields can be represented as
\begin{equation}
 \left\{\begin{array}{c}
         \Pi_0 \\ \Pi_\parallel \\ \Pi_\perp \\ \Pi_Q \\
        \end{array}\right\}
 =\frac{\alpha}{2\pi}\int_0^\infty\frac{{\rm d}s}{s}\int_{-1}^1\frac{{\rm d}\nu}{2}\left[{\rm e}^{-{\rm i}\Phi_0s}\,\frac{zz'}{\sin z\sinh z'}
 \left\{\begin{array}{c}
         k^2N_0 \\ v_\perp^2N_0-v_\parallel^2 N_1 \\ v_\perp^2 N_2-v_\parallel^2 N_0 \\ -N_3
        \end{array}\right\}
 +{\rm c.t.}\right], \label{eq:PII}
\end{equation}
with ${\rm c.t.}=-{\rm e}^{-{\rm i}m^2s}k^2(1-\nu^2)$ and
\begin{align}
 \Phi_0&=m^2-\frac{v_\perp^2}{2}\frac{\cos z-\cos(\nu z)}{z\sin z}-\frac{v_\parallel^2}{2}\frac{\cosh z'-\cosh(\nu z')}{z'\sinh z'}\,, \\
 N_0&=\cos(\nu z)\cosh(\nu z')-\cot z\sin(\nu z)\coth z'\sinh(\nu z')\,, \\
 N_1&=2\cos z\,\frac{\cosh z'-\cosh(\nu z')}{\sinh^2 z'}\,, \\
 N_2&=-2\cosh z'\,\frac{\cos z-\cos(\nu z)}{\sin^2 z}\,, \\
 N_3&=\frac{1-\cos z\cos(\nu z)}{\sin z}\frac{1-\cosh z'\cosh(\nu z')}{\sinh z'}+\sin(\nu z)\sinh(\nu z')\,,
\end{align}
where we employed the shorthand notations $z:=ec_+s$ and $z':=ec_-s$.
Moreover, the prescription $m^2\to m^2-{\rm i}0^+$ for the electron mass squared is implicitly understood here, and the propertime integration contour is assumed to lie slightly below the real positive $s$ axis

A straightforward but somewhat tedious calculation shows that
\begin{equation}
 F\partial_F v_{\parallel/\perp}^\mu=0\,,\quad{}^*\!F\partial_F v_{\parallel/\perp}^\mu=0 \,, \label{eq:FdFv}
\end{equation}
for ${\cal G}\geq0$, and
\begin{equation}
 F\partial_F u^\mu
 =2u^\mu\,,\quad {}^*\! F\partial_F u^\mu=0\,.
\end{equation}
With the help of these identities we infer
\begin{equation}
 F\partial_F P_{0,\parallel,\perp}^{\mu\nu}={}^*\!F\partial_F P_{0,\parallel,\perp}^{\mu\nu}=F\partial_F Q^{\mu\nu}={}^*\!F\partial_F Q^{\mu\nu}=0 \,,
\end{equation}
which implies that \Eqref{eq:Pi2} is spanned by the same tensor structures as the one-loop photon polarization tensor.
In turn, \Eqref{eq:Pi2} becomes
\begin{equation}
  \Pi^{\mu\nu}_{2\text{-loop}}(k|F)\big|_{1{\rm PR}}
 = \Delta\Pi_0 P_0^{\mu\nu}+\Delta\Pi_\parallel P_\parallel^{\mu\nu} +\Delta\Pi_\perp P_\perp^{\mu\nu} + \Delta\Pi_Q Q^{\mu\nu} \,,
 \label{eq:Pi3}
\end{equation}
where we defined 
\begin{equation}
 \Delta\Pi_{\cal I}
 = \frac{1}{2}\Bigl( \frac{\partial {\cal L}_{\text{HE}}^{1\text{-loop}}}{\partial{\cal F}}F\partial_F
 +\frac{\partial {\cal L}_{\text{HE}}^{1\text{-loop}}}{\partial{\cal G}}{}^*\!F\partial_F \Bigr)\Pi_{\cal I} \,.
 \label{eq:tildeM0}
\end{equation}
Finally, we can make use of the fact that $\Pi_{\cal I}$ depends on $z_k$ via $v_{\parallel}^2$ and $v_{\perp}^2$ only \cite{Dittrich:2000zu} and \Eqref{eq:FdFv} to express \Eqref{eq:tildeM0} as
\begin{equation}
 \Delta\Pi_{\cal I}
 = \biggl({\cal F}\frac{\partial {\cal L}_{\text{HE}}^{1\text{-loop}}}{\partial{\cal F}}
 +{\cal G}\frac{\partial {\cal L}_{\text{HE}}^{1\text{-loop}}}{\partial{\cal G}}\biggr)\frac{\partial\Pi_{\cal I}}{\partial{\cal F}}
 + \biggl({\cal G}\frac{\partial {\cal L}_{\text{HE}}^{1\text{-loop}}}{\partial{\cal F}}
  - {\cal F}\frac{\partial {\cal L}_{\text{HE}}^{1\text{-loop}}}{\partial{\cal G}}\biggr)\frac{\partial\Pi_{\cal I}}{\partial{\cal G}} \,,
 \label{eq:tildeM}
\end{equation}
with $\Pi_{\cal I}=\Pi_{\cal I}({\cal F},{\cal G},v_{\parallel}^2,v_{\perp}^2)$.

\subsection{Purely magnetic field}\label{subsec:2loopPi2}

For a purely magnetic field, \Eqref{eq:tildeM} reduces to
\begin{equation}
 \Delta\Pi_{\cal I}
 = \frac{1}{2}(\partial_B {\cal L}_{\text{HE}}^{1\text{-loop}})(\partial_B\Pi_{\cal I})\,,
 \label{eq:tildeM2}
\end{equation}
where we made use of the fact that $\Pi_{\cal I}$ is regular for ${\cal G}\to 0$ and ${\cal F}\to\frac{1}{2}B^2$, with $B=|\vec{B}|$ and $\partial_B=\frac{\partial}{\partial B}$.
As $\Pi_Q$ vanishes for ${\cal G}=0$ \cite{Dittrich:2000zu}, we moreover have $\Delta\Pi_0=0$.
Hence, the 1PR contribution to the photon polarization tensor in a magnetic field can be expressed as
\begin{equation}
  \Pi^{\mu\nu}_{2\text{-loop}}(k|B)\big|_{1{\rm PR}}
 = \frac{1}{2}(\partial_B {\cal L}_{\text{HE}}^{1\text{-loop}})\bigl[(\partial_B{\Pi_0}) P_0^{\mu\nu}+(\partial_B{\Pi_\parallel}) P_\parallel^{\mu\nu} +(\partial_B{\Pi_\perp}) P_\perp^{\mu\nu} \bigr] .
 \label{eq:Pi4}
\end{equation}
In a purely magnetic field it is moreover convenient to decompose four-vectors with regard to the direction of the applied magnetic field $\hat{\vec{B}}$ as 
\begin{equation}
  k^\mu=k^\mu_\parallel + k^\mu_\perp\,,\quad k_\parallel^\mu=(k^0,\vec{k}_\parallel)\,,\quad k_\perp^\mu=(0,\vec{k}_\perp)\,,
\end{equation}
with $\vec{k}_\parallel=(\vec{k}\cdot\hat{\vec{B}})\hat{\vec{B}}$ and $\vec{k}_\perp=\vec{k}-\vec{k}_\parallel$, and analogously $g^{\mu\nu}=g^{\mu\nu}_\parallel+g^{\mu\nu}_\perp$ \cite{Dittrich:2000zu}.
With these definitions, the projectors $P^{\mu\nu}_{0,\parallel,\perp}$ in \Eqref{eq:Pi4} can be written as
\begin{equation}
 P_\parallel^{\mu\nu}=g^{\mu\nu}_\parallel-\frac{k_\parallel^\mu k_\parallel^\nu}{k_\parallel^2}\,, \quad P^{\mu\nu}_\perp=g^{\mu\nu}_\perp-\frac{k_\perp^\mu k_\perp^\nu}{k_\perp^2}\,,
\end{equation}
and
\begin{equation}
 P_0^{\mu\nu}=\Bigl(g^{\mu\nu}-\frac{k^\mu k^\nu}{k^2}\Bigr)-P_\parallel^{\mu\nu}-P_\perp^{\mu\nu}\,.
\end{equation}

A compact parameter integral representation of $\partial_B{\cal L}_{\text{HE}}^{1\text{-loop}}$ can be inferred from the first line of Eq.~(11) of Ref.~\cite{Karbstein:2015cpa} and is given by\footnote{Note that many superficially different representations of \Eqref{eq:dLdB} exist, which are related by partial integrations with respect to the propertime variable $z$.}
\begin{equation}
 \partial_B {\cal L}_{\text{HE}}^{1\text{-loop}}
 =B\frac{\alpha}{2\pi}\int_{0}^{\infty}\frac{{\rm d}z}{z}\,{\rm e}^{-{\rm i}\frac{m^2}{eB}z}\biggl(\frac{\cot z}{z}-\frac{1}{\sin^2 z}+\frac{2}{3}\biggr) . \label{eq:dLdB}
\end{equation}
For \Eqref{eq:dLdB} the same implicit assumptions as given below \Eqref{eq:PII} apply.
For an explicit representation of~\Eqref{eq:dLdB} in terms of derivatives of the Hurwitz zeta function, cf. Ref.~\cite{Karbstein:2015cpa}.
Analogous parameter integral representations of the derivatives $\partial_B\Pi_{\cal I}$ entering \Eqref{eq:Pi4} follow straightforwardly from \Eqref{eq:PII} in the limit of $c_+\to B$ and $c_-\to0$.
More specifically, in order to perform the derivative for $B$ most efficiently, it is convenient to first perform the following substitution of the propertime variable: $s\to\frac{z}{eB}$, and only then perform the derivative for $B$.
This results in the following expression,
\begin{equation}
 \left\{\begin{array}{c}
         \partial_B\Pi_0 \\ \partial_B\Pi_\parallel \\ \partial_B\Pi_\perp 
        \end{array}\right\}
 =\frac{1}{B}\frac{\alpha}{2\pi}\left[\int_0^\infty{\rm d}z\int_{-1}^1\frac{{\rm d}\nu}{2}\,{\rm i}\frac{\Phi_0}{eB}\,{\rm e}^{-{\rm i}\frac{\Phi_0}{eB}z}\,\frac{z}{\sin z}
 \left\{\begin{array}{c}
         k^2N_0 \\ k_\perp^2N_0+k_\parallel^2 N_1 \\ k_\perp^2 N_2+k_\parallel^2 N_0
        \end{array}\right\}-\frac{2}{3} k^2\right] , \label{eq:dPiBdB}
\end{equation}
where now
\begin{align}
 \Phi_0&=m^2+k_\parallel^2\frac{1-\nu^2}{4}+k_\perp^2\frac{\cos\nu z-\cos z}{2z\sin z}\,, \nonumber\\
 N_0&=\cos\nu z-\nu\sin\nu z\cot z\,, \nonumber\\
 N_1&=(1-\nu^2)\cos z\,, \nonumber\\
 N_2&=2 \frac{\cos \nu z-\cos z}{\sin^2 z}\,.
\end{align}

Aiming at analytical insights into $\Pi^{\mu\nu}_{2\text{-loop}}(k|B)\big|_{1{\rm PR}}$ in a given parameter limit, instead of manipulating the exact expressions given in Eqs.~\eqref{eq:dLdB} and \eqref{eq:dPiBdB}, 
it is typically more convenient to turn to the known expressions for ${\cal L}_{\text{HE}}^{1\text{-loop}}$ and $\Pi_{\cal I}$ in this limit and perform the derivatives of $B$ thereof.
This immediately implies that analytical results for $\Pi^{\mu\nu}_{2\text{-loop}}(k|B)\big|_{1{\rm PR}}$ are available in the same cases as for $\Pi^{\mu\nu}_{1\text{-loop}}(k|B)$.
A detailed analysis of $\Pi^{\mu\nu}_{1\text{-loop}}(k|B)$, putting special emphasis on the various limits where analytical insights are available can be found in Ref.~\cite{Karbstein:2013ufa}.

Let us also recall that due to an electric-magnetic duality, the Heisenberg-Euler effective Lagrangian and the photon polarization tensor in a purely magnetic and a purely electric field are related and can be translated into each other \cite{Jentschura:2001qr,Karbstein:2013ufa}: The magnetic field result is converted into the corresponding electric field result by means of the transformations $B\leftrightarrow -{\rm i}E$ and $\parallel\leftrightarrow \perp$. 
Hence, our result~\eqref{eq:Pi4} for $\Pi^{\mu\nu}_{2\text{-loop}}(k|B)\big|_{1{\rm PR}}$ can be straightforwardly adapted to a purely electric field, yielding the exact expression for $\Pi^{\mu\nu}_{2\text{-loop}}(k|E)\big|_{1{\rm PR}}$.

\section{Conclusions and Outlook}\label{sec:CaO}

In this article, we have outlined how a given 1PI contribution evaluated in constant external electromagnetic fields, accounting for the coupling of the external field to the charged particle propagators to all orders,
can be employed to efficiently determine 1PR tadpole contributions to the same quantity.
More specifically, we have demonstrated that all 1PR tadpole contributions to a given quantity follow by differentiations of lower loop diagrams for the field strength tensor and contractions with derivatives of the 1PI contribution to the constant-field Heisenberg-Euler effective Lagrangian.
This in particular implies that all 1PR tadpole diagrams to a given quantity in constant fields are fully determined by 1PI diagrams.
Given that the generating 1PI diagrams are known explicitly, along these lines the evaluation of the corresponding 1PR tadpole diagrams becomes trivial.

As an application of our approach, we have explicitly determined the 1PR contribution to the two-loop photon polarization tensor for QED in constant electromagnetic fields.
Together with the corresponding, not yet determined 1PI contribution at two loops, this will constitute the full expression of the two-loop photon polarization tensor for external-field QED in $d=3+1$ dimensions.
Particularly for a purely magnetic or a purely electric field, our result can be cast into a rather compact expression, reminiscent of the corresponding one-loop photon polarization tensor.

We expect our results to be relevant for precise theoretical studies of various QED processes in constant electromagnetic fields at higher loops.

\acknowledgments

It is a pleasure to thank Holger Gies for many stimulating discussions and valuable comments on this manuscript

\end{document}